\documentclass[aps,prl,twocolumn,amssymb,longbibliography,amsmath,superscriptaddress]{revtex4-2}

\usepackage[utf8]{inputenc}
\usepackage[colorlinks=true]{hyperref}

\usepackage{amsmath}
\usepackage{amssymb}
\usepackage{amsfonts}
\usepackage{graphicx}
\usepackage{verbatim}
\usepackage{bm}
\usepackage{mathbbol}
\usepackage{color}
\setcitestyle{compress}

\usepackage{mathtools}

\usepackage{graphicx,epsfig,amsfonts,amssymb}
\usepackage{bm}
\usepackage{lipsum}
\usepackage{verbatim}
\usepackage{ulem}

\usepackage{hyperref}
\hypersetup{
    colorlinks,
    citecolor=blue,
    filecolor=blue,
    linkcolor=blue,
    urlcolor=blue
}

\makeatletter
  \long\def\pprintMaketitle{\clearpage
  \iflongmktitle\if@twocolumn\let\columnwidth=\textwidth\fi\fi
  \resetTitleCounters
  \def\baselinestretch{1}%
  \printFirstPageNotes
  \begin{center}%
 \thispagestyle{pprintTitle}%
   \def\baselinestretch{1}%
    \Large\@title\par\vskip18pt
    \normalsize\elsauthors\par\vskip10pt
    \footnotesize\itshape\elsaddress\par\vskip36pt
    \end{center}%
  \gdef\thefootnote{\arabic{footnote}}%
  }
\makeatother

\makeatletter
\def\ps@pprintTitle{%
  \let\@oddhead\@empty
  \let\@evenhead\@empty
  \def\@oddfoot{\reset@font\hfil\thepage\hfil}
  \let\@evenfoot\@oddfoot
}
\makeatother

\newcommand{\beg}{\begin{equation}}
\newcommand{\en}{\end{equation}}

\usepackage{soul}
 
\begin{document}

\title{Rydberg dressed spin-1/2 Fermi gases in one dimension}

\author{Junhyun Lee}
\affiliation{Department of Physics and Astronomy, Center for Materials Theory, Rutgers University, Piscataway, NJ 08854 USA}

\author{Pavel A. Volkov}
\affiliation{Department of Physics and Astronomy, Center for Materials Theory, Rutgers University, Piscataway, NJ 08854 USA}
\affiliation{Department of Physics, Harvard University, Cambridge, Massachusetts 02138, USA}
\affiliation{Department of Physics, University of Connecticut, Storrs, Connecticut 06269, USA}

\author{B. J. DeSalvo}
\affiliation{Department of Physics, Indiana University, Bloomington, IN 47405 USA}
\affiliation{Quantum Science and Engineering Center, Indiana University, Bloomington, IN 47405 USA}
  
\author{J. H. Pixley}
\affiliation{Department of Physics and Astronomy, Center for Materials Theory, Rutgers University, Piscataway, NJ 08854 USA}
\affiliation{Center for Computational Quantum Physics, Flatiron Institute, New York, New York 10010, USA}

\begin{abstract}
    The emergent phases of strongly correlated spin-1/2 Fermi gases of Rydberg dressed atoms in a one dimensional optical lattice are theoretically investigated. 
    At weak coupling a bosonization description is used to demonstrate the ability to drive  alternating  quantum phase transitions between distinct Luttinger liquids. At strong coupling the ground state develops non-trivial phase separation exhibiting Luttinger liquid ``puddles'' separated by magnetic domain walls due to the interplay of the incommensurate filling and the Rydberg core length scale.  
    These phases can be detected  in ultracold gases of Rydberg atoms made from  $^6$Li.
\end{abstract}
\maketitle

\date{\today}

Over the last few decades, ultracold atoms have emerged as an ideal platform for quantum simulation. Offering extraordinary levels of experimental control including single-site imaging \cite{Gross2021} and tunable contact interactions \cite{Chin2010}, these systems offer a unique window into the world of quantum many-body physics. Recently, many theoretical and experimental efforts have focused on realizing long-range interactions in ultracold systems by using ultracold molecules \cite{DeMarco2019}, dipolar atoms \cite{Chomaz2022}, mediated interactions \cite{DeSalvo2019,Edri2020}, and Rydberg-dressed atoms \cite{Zeiher2016} in order to expand the scope of physics accessible in these systems. These efforts have yielded exciting advances realizing bound quantum droplets and supersolids \cite{Bottcher2020}. 

In contrast to other methods that induce long-range interactions between atoms, the scheme of Rydberg dressing offers the ability to locally tune interactions on fast timescales. Further, these interactions have been recently predicted to give rise to a host of exotic topological phases including $p$-wave superfluids \cite{Xiong2014}, topological Mott insulators \cite{Cardarelli2022}, and fractional Chern insulators \cite{Zhao2022}. Early experimental efforts to explore Rydberg dressed atoms focused on heavy bosonic atoms \cite{Balewski2014, Zeiher2016, Aman2016}, and found faster than expected loss rates consistent with an avalanche decay mechanism \cite{Aman2016, DeSalvo2016, Goldschmidt2016, Boulier2017}.  Experimentally, this loss processes can be mitigated by working with light fermionic atoms in low-dimensions, and recent experiments have shown promising results using $^6$Li in 2D \cite{Guardado-Sanchez2021}.  Based on these results, it is expected that one-dimensional (1D) fermionic gases will be the optimal experimental setting for Rydberg dressed atoms.

\begin{figure}[h!]
    \centering
    \includegraphics[width=0.47\linewidth]{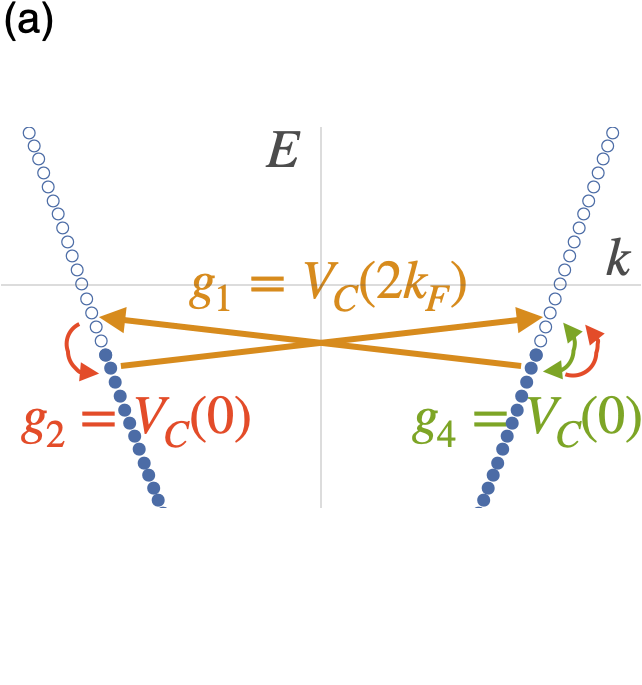}
    \includegraphics[width=0.51\linewidth]{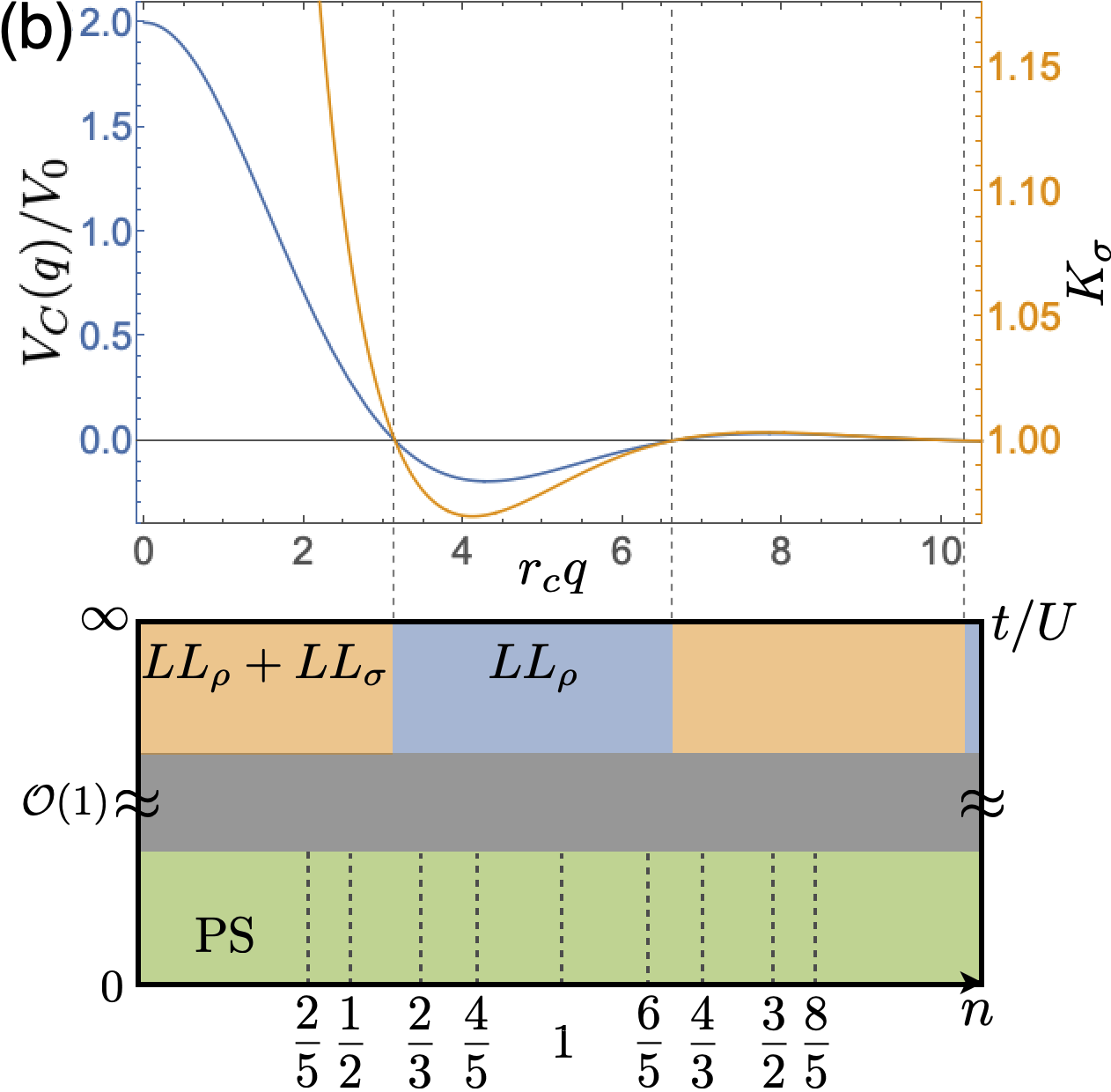}
    \caption{
    (a) A schematic figure of a 1D dispersion filled  (solid symbols) to the  Fermi momenta $k_F$ and the `$g$-ology' of the forward ($g_2,g_4$) and back ($g_1$) scattering near the Fermi-surface and its relation to the dressed Rydberg potential $V_C(q)$. 
    (b, Top) The dressed Rydberg interaction potential in momentum space and the continuum limit, $V_C(q)$, 
    plotted with the weak-coupling result for spin ($\sigma)$ Luttinger parameter $K_\sigma$, see Eq.\eqref{eq:weakksigma}. 
    (b, Bottom) Phase diagram of the system as a function of filling $n$ for weak ($t/U \gg 1$) and strong ($t/U \ll 1$) coupling. The weak coupling regime with $g_1>0$ is described by gapless charge ($\rho$) and spin Luttinger liquids (LL; denoted $LL_\rho + LL_\sigma$). For $g_1<0$ the spin sector is gapped out leaving only a charge LL ($LL_\rho$). At strong coupling, states with commensurate even and odd fillings are paramagnetic and magnetized, respectively. For incommesnurate fillings, the system is phase-separated (PS) into regions of commensurate density.
    }
    \label{fig:V}
\end{figure} 

Motivated by these  considerations we study a spin-1/2 Fermi gas of Rydberg dressed atoms in  a 1D optical lattice. We determine the emergent phases that the spin-1/2 fermion nature of the problem brings forth, in comparison to its bosonic counterpart. An intriguing feature of the dressed Rydberg interaction potential in 1D is that it is a oscillatory and sign changing function of momentum, which implies that the sign of the backscattering interaction can be controlled experimentally via the density or the core size of the Rydberg dressed potential, as shown in Fig.~\ref{fig:V}. As a result, we demonstrate that for incommensurate fillings in the weak coupling regime the ground state alternates back and forth as a function of the density (or the core size of the Rydberg dressed interaction) between a spin-charge separated Luttinger liquid and a  charge Luttinger liquid with a spin gap. In the strong coupling, but non-zero tunneling regime, we discover a series of ground states that are phase separated into ``puddles'' of Luttinger liquids separated by antiferromagnetic domain walls that act as an entanglement barrier.
The phase diagram in the weak and strong coupling regimes summarizing this behavior is shown in Fig.~\ref{fig:V}.

{\it Model and Approach}:
\label{sec:model}
We start  from a microscopic one-dimensional  model Hamiltonian for a spin-1/2 Fermi gas with a dressed Rydberg interaction in an optical lattice 
\begin{equation}
    H= -t\sum_{i,\sigma} (c_{i+1,\sigma}^{\dag}c_{i,\sigma} + \mathrm{h.c.}) + \sum_{i,j}V_d(|i-j|) n_i n_j.
    \label{eqn:HL}
\end{equation}
where $n_i=\sum_{\sigma}c_{i,\sigma}^{\dag}c_{i,\sigma}$.
$c_{i\sigma}^{\dag}$ is a creation operator of the  Rydberg dressed state at site $i$: $c_{i\sigma}^{\dag}|i;0\rangle=|i;g,\sigma\rangle +\beta|i;R,\sigma\rangle$ with spin $\sigma=\uparrow,\downarrow$. The state is weakly mixed between the ground ($g$) and the Rydberg level ($R$) with $\beta=\Omega/(2\Delta)\ll 1$ where $\Omega$ is the Rabi frequency between the ground and Rydberg levels, and $\Delta$ is the detuning. The dressed Rydberg interaction potential $V_d$ is

\begin{equation}
    V_d(|i-j|)=\frac{U}{r_c^6+|i-j|^6},
    \label{eq:rydbergint}
\end{equation}
where the coupling strength  $U \equiv C_6 \beta^4$, $C_6$ is the van der Waals coefficient,
and the ``core'' of the interaction is $r_c \equiv (C_6 /2 \Delta)^{1/6}$~\cite{Honer10,Balewski2014,Zeiher2016}.

We study the model defined in Eq.~\eqref{eqn:HL} using a combination of field theoretic bosonization techniques in the weak coupling regime, numerical simulations using the density matrix renormalization group (DMRG) approach, and classical simulations of the density profile that are helpful in interpreting the strong coupling regime. 
We utilize the DMRG method to compute the ground state of the lattice model (with lattice spacing $a=1$) in  Eq.~\eqref{eqn:HL} in a one-dimensional chain with open boundaries \footnote{See Supplemental Material for the effects of a harmonic trap in DMRG, discussion of density correlation functions and bosonization arguments in the phase separated regime and details of PXP limit and weak coupling calculations}.
The largest system size we have considered is $L=400$, and we have set the maximum bond dimension as $1200$ to keep the truncation error to be below $10^{-12}$ for most calculations. (for weak coupling and $N_p=120$, the truncation error was about $10^{-8}$ with bond dimension $2000$) 
All DMRG calculations in this paper are performed using the ITensor Library~\cite{itensor}.

{\it Weak coupling limit}:
\label{sec:weak}
In the weak coupling regime of the Rydberg dressed Hamiltonian [Eq.~\eqref{eqn:HL}], the low energy field theoretic description of the problem is obtained by linearizing the dispersion near the Fermi momenta $\pm k_F$ [Fig. \ref{fig:V} (a)].
The interaction in Eq.~\eqref{eq:rydbergint} in momentum space takes the form $\frac{1}{2\pi L}\int dq V_C(q) n(q) n(-q)$. 
In the limit of $r_c\gg a$,
$V_C(q)$ is equal to the Fourier transform of the continuum limit of Eq.~\eqref{eq:rydbergint} \cite{Note1}:
\begin{equation}
    V_C(q) = 
    V_0 \left[e^{-|q| r_c} 
    + 2e^{-\frac{|q| r_c}{2}}\cos \left( 
    |q| r_c\sqrt{3}/
    2 - \phi\right) \right],
    \label{eqn:VC}
\end{equation}
where $\frac{\pi U}{3 r^5_c}  \equiv V_0$,  $\phi = \pi/3$.
An important feature of $V_C(q)$, plotted in Fig.~\ref{fig:V}(b), is that it changes sign from a repulsive to an attractive 
interaction as a function of $|q|$. 

Applying the standard `$g$-ology' of relevant interaction induced scattering process in 1D, leads us naturally to consider 3 distinct scattering channels~\cite{Giamarchi}, depicted in Fig.~\ref{fig:V}(a).
The $g_4$($g_2$)-process corresponds to forward scattering between the fermions on the same(different) side(s) of the Brillouin zone and thus $g_2, g_{4} \propto V_C(0)$.
In contrast, the $g_1$-process scatters two fermions from the left/right side to the right/left side and therefore dubbed as backscattering. The momentum transfer is $2k_F$ in this case, and the coupling constant is $g_1 \propto V_C(2k_F)$. Due to the spin rotation invariance, the interaction in the parallel and transverse spin channels have the same strength, i.e. $g_{i,\parallel} = g_{i,\perp}$.
Taking Eq.~\eqref{eqn:VC} into account directly shows that by tuning the number density $\rho=2k_F a/\pi$
of the fermions we are able to control the sign of the effective backscattering interaction and change it from repulsive to attractive, and vice versa.

In the weak coupling regime, we can use abelian bosonization to describe the low-energy physics of the ground state.
This leads to the standard spin-charge separated Hamiltonian 
in terms of a Luttinger liquid away from commensurate fillings:
$H^0_\eta = \frac{1}{2\pi} \int dx \left[ u_\eta K_\eta (\pi \Pi_\eta (x))^2 + \frac{u_\eta}{K_\eta} (\nabla \phi_\eta(x))^2 \right]$ where $\eta=\rho,\sigma$ corresponds to charge and spin degrees of freedom, respectively, and the spin sector has the interaction
\begin{align}
    H_\sigma =& H_\sigma^0
    +\frac{2 V_C(2k_F)}{(2\pi \alpha)^2} \int dx \cos(\sqrt{8} \phi_\sigma),
    \label{eqn:cos}
\end{align}
where $\alpha$ is the short-distance (lattice) cutoff scale. The charge part ($H_\rho^0$) is the Luttinger liquid Hamiltonian, while the spin part ($H_\sigma$) is the sine-Gordon Hamiltonian.
$\phi_{\rho(\sigma)}$'s are the bosonized fields for the charge (spin) sector and $\Pi_{\rho(\sigma)}$'s are its conjugate momentum.
$u_\eta$'s and $K_\eta$'s are the  velocities and Luttinger parameters for the corresponding sectors $\eta=\rho,\sigma$.
The relevance of the cosine term in Eq.~\eqref{eqn:cos} is dictated by the value of the Luttinger parameter in the spin sector,  that obeys~\cite{Giamarchi}:
\begin{align}
    K_\sigma &= 
    \left[\frac{1 + V_C(2k_F)/2 \pi v_F}{1 - V_C(2k_F)/2 \pi v_F}\right]^{1/2},
   \label{eq:weakksigma}
\end{align}
plotted in Fig.~\ref{fig:V}, with $r_c = 4a$,  $v_F=(2t/a)\sin(a k_F)$, and $V_0/t= 1$. When $V_C(2k_F)$ changes sign from positive to negative the cosine interaction switches from irrelevant ($K_\sigma>1$) to relevant ($K_\sigma<1$). In the latter case, the interaction \eqref{eqn:cos} ``locks'' the field to the value $\phi_\sigma = 2n\pi/\sqrt{8}$ and excitations of $\phi_\sigma$ become gapped. The spin-density wave order parameter in the $z$-direction $\mathcal{O}^z_{\textrm{SDW}} = \psi^\dagger_{R\uparrow}\psi_{L\uparrow} - \psi^\dagger_{R\downarrow}\psi_{L\downarrow}\sim \sin \sqrt{2} \phi_\sigma$ 
vanishes, and its correlations $\langle \mathcal{O}^{z \dagger}_{\textrm{SDW}}(x)\mathcal{O}^z_{\textrm{SDW}}(0)\rangle$ decay exponentially. 
(The $L/R$ subscript corresponds to the left/right moving fermionic field). The charge correlations remain gapless in this phase and behave as a power-law, in particular charge-density wave correlation decay as $\sim x^{-K_\rho}$ while superconducting ones as $\sim x^{-K_\rho^{-1}}$. As $K_\rho^2 =(1+ V_C(2k_F)/2 \pi v_F)/(1+2 V_C(0)/ \pi v_F- V_C(2k_F)/2 \pi v_F)<1$,
the CDW correlations are dominant.

This is in contrast with when backscattering is repulsive; $g_1=V_C(2k_F)>0$ and $K_\sigma>1$, and both spin and charge sectors are gapless. 
The SDW correlation function decays as a power-law  $\sim x^{-(K_\rho + K_\sigma)}$. Moreover, as $K_\rho<1$ also in this case, CDW and SDW correlations dominate over the superconducting ones \cite{Giamarchi}.

Therefore, labeling the momentum transfer where $V_C(q)$ first becomes negative to be $q_-$, and $q_+$ when is becomes again positive implies that $q_\pm = 2 k_F$ would correspond to a transition between a charge and spin Luttinger liquid and a spin-gapped charge Luttinger liquid, which belongs to Berezinskii–Kosterlitz–Thouless (BKT) class \cite{Giamarchi}. Note that the sign-change of $V_C(q)$ is a function of $q r_c$, and thus the phase transition can be tuned by either density (through $k_F$) or $r_c$.

\begin{figure}[t!]
    \centering
    \includegraphics[width=0.49\linewidth]{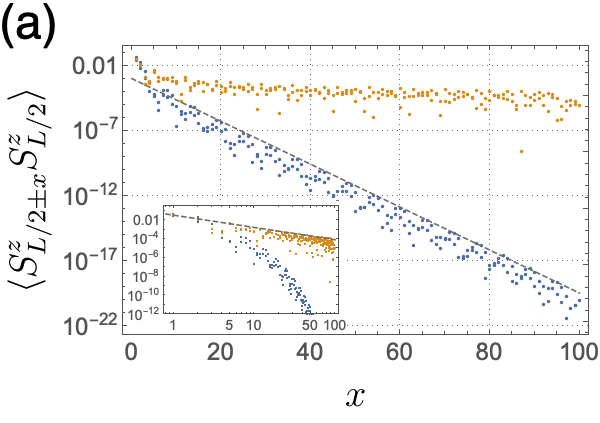}
    \includegraphics[width=0.49\linewidth]{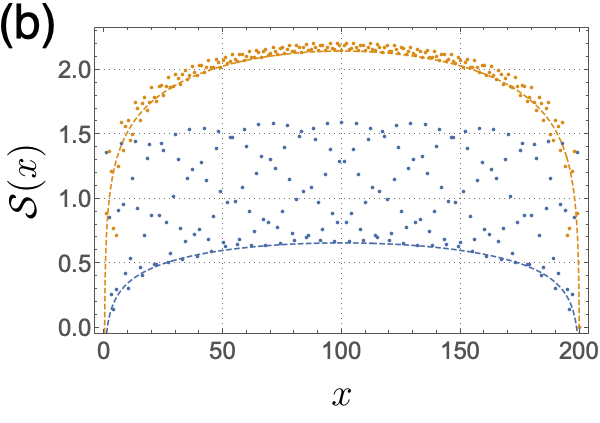}
    \caption{(a) The spin-spin correlation function for $N_p = 70$ (blue) and $N_p = 120$ (yellow) particles in a $L = 200$ chain with $U/ r_c^6 = 2t$, $r_c = 4a$.
    $x=0$ is defined as the site $L/2$, and the data include both sides of $x=0$.
    The backscattering is attractive when $N_p = 70$, and repulsive when $N_p = 120$. 
    The correlation shows a clear exponential decay when the backscattering is attractive. 
    The gray dashed lines are the guide to the eye following the envelope function.
    (b) The bipartite entanglement entropy as a function of the cut position. 
    The dashed lines are the entanglement entropy scaling corresponding to $c = 1$ (blue $N_p=70$) and $c=2$ (yellow $N_p=120$).
    }
    \label{fig:weak}
\end{figure} 

We now verify these predictions numerically by calculating the ground state of a $L=200$ chain with various fillings. 
We fix the $U/ r_c^6 = 2t$, $r_c = 4a$ and tune the number of fermions to go across the density-tuned phase transition.
The particle number corresponding to $q_-$ and $q_+$ is $N_p \approx 50$ and $N_p \approx 106$, respectively, for the $L$ and $r_c$ we consider. 
In Fig.~\ref{fig:weak} we choose two representative fillings, $N_p = 70$ and $N_p = 120$ which are both deep within each phase and plot their spin-spin correlation function in Fig.~\ref{fig:weak}(a). \cite{Note1}
We observe that the spin-spin correlations decays exponentially for $N_p = 70$, and as a power-law for $N_p = 120$. 

We complement the correlation function analysis with another observable that indicates the gapped spin sector, the von-Neumann entanglement entropy. 
We extract the central charge, which counts the number of gapless modes in the system, from the scaling behavior of the bipartite entanglement entropy $\mathcal{S}$ as we tune the cut position $x$~\cite{CalabreseCardy2004},
\begin{align}
    \mathcal{S}(x) = \frac{c}{6}\ln \left[\frac{L}{\pi}\sin\left(\frac{\pi x}{L}\right)\right] + d,
    \label{eq:ee}
\end{align}
where $c$ is the central charge of the system and $d$ is a nonuniversal constant including the boundary entropy~\cite{affleck1991}.
This scaling behavior is shown in Fig.~\ref{fig:weak}(b), for the same system: $L=200$, $r_c=4a$, and $N_p = 70, 120$.
With the dashed lines corresponding to $c = 1$ and $c = 2$ of Eq.~\eqref{eq:ee}, we confirm that the two fillings belong to the $c=1$ phase of charge Luttinger liquid (spin-gapped), and $c=2$ phase of charge and spin Luttinger liquid, respectively.

\begin{figure}[t!]
    \centering
    \includegraphics[width=0.8\linewidth]{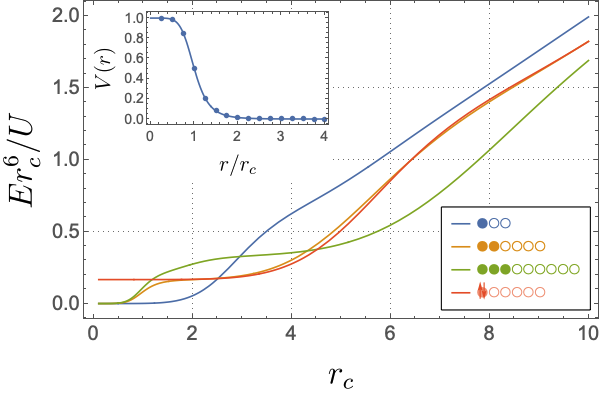}
    \caption{Normalized energy per site as a function of $r_c$ for different filling patterns with same average density $\rho = 1/3$ computed in the classical limit ($t=0$). 
    Note that the arrows in the pattern of the red line indicates the up and down spin, and thus have a unit cell of six sites with one doubly occupied site. 
    }
    \label{fig:classical}
\end{figure} 

{\it Strong coupling limit}:
\label{sec:strong}
\begin{figure*}[t!]
    \centering
    \includegraphics[width=0.32\linewidth]{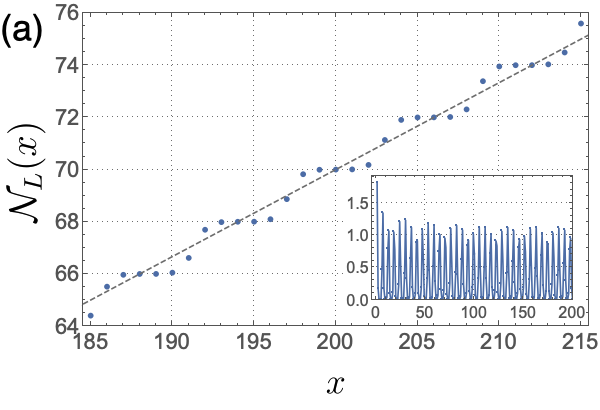}
    \includegraphics[width=0.32\linewidth]{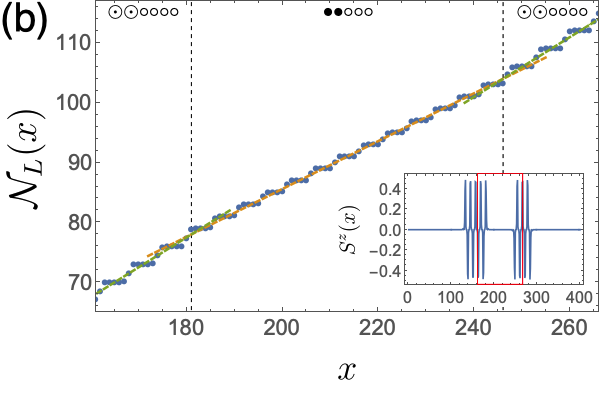}
    \includegraphics[width=0.32\linewidth]{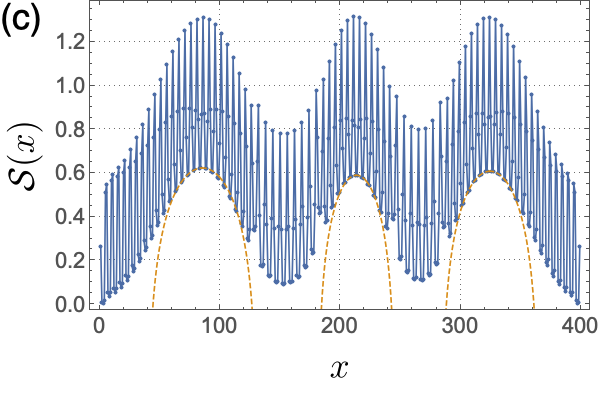}
    \caption{Left accumulated particle $\mathcal{N}_L(x)$ as a function of position for $N_p = 140$ [(a)] and $N_p = 170$ [(b)] particles in a $L = 400$ chain.
    (a) $\bullet$%
    $\bullet$%
    $\circ$%
    $\circ$%
    $\circ$%
    $\circ$ pattern is observed, and the dashed line with slope $1/3$ follows the average data well. 
    The inset shows the particle density $n(x)$ up to the center of the chain, which shows there are above average particles accumulated at the boundary. 
    (b) Demonstrating phase separation  as the $\bullet$$\bullet$$\circ$$\circ$$\circ$ pattern is observed at both edges and the center of the chain, forming three commensurate regions with density $\rho = 2/5$, shown by the slope of the yellow dashed line (here we have zoomed into the center of the chain). These are separated by two domains with a  
    $\odot$%
    $\odot$%
    $\circ$%
    $\circ$%
    $\circ$%
    $\circ$ pattern that have   two occupied sites with 1.5 particles each and the slope of the green dashed line is $1/2$.
    The inset shows the spin density $S^z(x)=(n_\uparrow(x) - n_\downarrow(x))/2$ of the system, demonstrating the  region with 
    $\odot$%
    $\odot$%
    $\circ$%
    $\circ$%
    $\circ$%
    $\circ$ 
    has a non-zero staggered magnetization. 
    (c) The bipartite entanglement entropy as a function of the cut position for the phase separated case $N_p = 170$. The three dashed lines corresponds to $c = 1$ of Eq.~\eqref{eq:ee} for each partition. 
    }
    \label{fig:strong}
\end{figure*}  
In the strong coupling limit $U\gg t$ we find that commensurability effects start playing a major role. Even at nominally incommensurate densities the system separates into regions that have commensurate density as is shown schematically in Fig.~\ref{fig:V} (b). We find that neighboring pairs of commensurate densities with an odd or even number of sites per unit cell split into two types of phases that are either magnetized or not, respectively allowing  phase separation to become energetically favorable.

At commensurate fillings (or in the interiors of phase-separated regions), one may expect a periodic density-ordered patter to emerge \cite{Giamarchi,schulz_mott}. This pattern 
depends on the average density, but also $r_c$. This can be straightforwardly seen in the `classical limit' (setting $t=0$ in Eq.~\eqref{eqn:HL}).
Fig.~\ref{fig:classical} shows the average classical energy per site, 
for different filling patterns with the same average density. 
One can observe that the energy minimum changes as we tune $r_c$. 
In general, the optimal unit cell size increases with $r_c$.

We now discuss DMRG results on the full quantum system Eq.~\eqref{eqn:HL} in the strong coupling limit: $U/ r_c^6 = 10t$ and $r_c = 4a$. Fig.~\ref{fig:strong}(a) shows the DMRG calculation for an $L=400$ site chain 
where the bulk has the commensurate density $\rho=1/3$ 
(dashed line in the phase diagram of Fig.~\ref{fig:V}(b)).
Here, the $y$-axis is defined as $\mathcal{N}_L(x) = \sum^x_{i=1} \langle n_i \rangle$ which is the expectation value of accumulated particle number to the left of position $x$.
For example, if the particle was evenly distributed across the system with a density $\rho_0$, then the $\mathcal{N}_L(x) - x$ plot will show a straight line of slope $\rho_0$.
From the figure, it is clear that $\mathcal{N}_L(x)$ increases by $2$ every six sites, that is, in an average density of $1/3$. 
More precisely, the pattern shows that the increase of $\mathcal{N}_L(x)$ occurs mostly at two adjacent sites, and changes little on the next four sites. 
This is roughly a unit cell of six sites with repeating pattern, $\bullet$$\bullet$$\circ$$\circ$$\circ$$\circ$, 
where the open(filled) circles are empty(single occupied) sites. In the classical calculation the ground state of $\rho = 1/3$ and $r_c = 4a$ is the doubly occupied pattern (red line in Fig.~\ref{fig:classical}) and the $\bullet$$\bullet$$\circ$$\circ$$\circ$$\circ$ pattern has the second lowest classical energy. This shows that even for $U/t=10$, quantum effects are important to determine the ground state. Moreover, the correlations and entanglement properties of this state \cite{Note1} indicate that the system does not have a charge gap despite the strong density modulation.

The number of particles in this calculation is $N_p = 140$, and this ratio approaches the bulk density in the thermodynamic limit, i.e. $\rho=N_p / L \rightarrow 1/3$ as  $L \rightarrow \infty$, defining the bulk phase.
The open boundaries introduces a boundary effect such that the density at the edge is different from that of the bulk (see inset of Fig.~\ref{fig:strong}(a)). 
While this effect will vanish in systems with periodic boundary conditions, we have checked that this ground state phase is stable to the experimentally relevant inclusion of a harmonic trap\cite{Note1}.

Fig.~\ref{fig:strong}(b) is the same calculation but with $N_p = 170$ particles and its entanglement entropy  is shown in Fig.~\ref{fig:strong}(c), representative of one of  the regions of
incommensurate filling in the phase diagram in Fig.~\ref{fig:V}(b). 
Distinct, effectively commensurate regions are observed through their different average slopes of $\mathcal{N}_L(x) - x$ versus $x$.
At the edges and the center of the the system [see Fig.~\ref{fig:strong}(b, inset) and ~\ref{fig:strong}(c) for the full system] $\mathcal{N}_L(x) - x$ has a slope of $2/5$ (which is depicted as an orange dashed line) with a pattern of 
$\bullet$%
$\bullet$%
$\circ$%
$\circ$%
$\circ$. 
Between the $2/5$ commensurate regions, there are two regions with a
$\odot$%
$\odot$%
$\circ$%
$\circ$%
$\circ$%
$\circ$ pattern and a slope equal to $1/2$  (depicted as a green dashed line), but with three particles occupying the two filled circles (the $\odot$ symbol indicates a site with an occupation of 3/2  particles on average).
This example demonstrates the phase separation in the strong coupling regime: placing particles at the boundary is not enough and the system phase separates into a simple fraction with two fillings. Indeed, the penalty of excess particles at the boundary depends on the interaction strength $U/t$ and the long-range interaction, and the system will be more likely to phase separate in stronger interactions. Last, we show that this phase separation survives the inclusion of a harmonic trap~\cite{Note1}.

The magnetization in this case (the inset of Fig.~\ref{fig:strong}(b)) is staggered for the $\odot$$\odot$$\circ$$\circ$$\circ$$\circ$ regions but zero outside of it.
This is consistent with the expectation that an even number of spins per unit cell will form local singlets, while odd number allows for gapless spin correlations.
Another interesting feature of this regime is that each region retains its properties as a Luttinger liquid ``puddle.'' 
For instance,
the charge excitations are gapless in the $\bullet$$\bullet$$\circ$$\circ$$\circ$ phase, while gapped in the magnetized $\odot$$\odot$$\circ$$\circ$$\circ$$\circ$ phase. This can be seen in Fig.~\ref{fig:strong}(c) where we plot the bipartite entanglement entropy of the system.
The three regions of $\bullet$$\bullet$$\circ$$\circ$$\circ$ phase are nicely fitted with the entanglement scaling [Eq.~\eqref{eq:ee}] with $c=1$, which comes from the gapless charge degrees of freedom. 
The magnetized region also serves as a barrier for information spreading~\cite{Note1}. 

This behavior can be qualitatively understood from general bosonization arguments. The phases presented in Fig. \ref{fig:strong} have band filling of $1/m \equiv \rho/2 $ with $m=6$ (a), $m=5$ (b,c puddles with $c\approx 1$) and $m=4$ (b,c, magnetized regions). For these fillings, a charge gap is expected to open \cite{schulz_mott,Giamarchi} for $K_\rho<4/m^2$  \cite{Note1}, requiring quite strong interactions. The observation of gapless $2k_F$ density correlations for $m=5,6$, but not $m=4$ suggests that $0.16<K_\rho<0.25$. For all three $m$ values $V_C(2k_F)/V_0<0$, such that a spin gap may be expected at weak coupling, but its value is very small for $m=4$ and $V_C(2k_F)/V_0\approx-0.026$. This provides an explanation for the apparent spin ordering for $m = 4$ in the DMRG results, rather then the uniform spin-gapped state one might expect from a weak coupling perspective.

{\it Optical Tweezers}: We now demonstrate that fermionic Rydberg atoms in optical tweezers can also be mapped to a so-called $PXP$ model  in the strong Rydberg coupling $\beta\gg1$ limit. For the filling of two atoms per site, the laser will couple the state where both are in ground state $|g\uparrow, g\downarrow\rangle$ to a single state $(|R\uparrow,g\downarrow\rangle + |g\uparrow,R\downarrow\rangle)/\sqrt{2}$, realizing the usual $PXP$ model~\cite{Turner18,turner2018}. On the other hand, for a single fermion per site, the spin degree of freedom is decoupled from light for $s$-like states, realizing a $PXP$ model with an additional spin degree of freedom per site, that can be manipulated, e.g. by a magnetic field. Spin exchange interactions may lift this degeneracy, but are likely to be negligible in current setups \cite{Note1}.

{\it Conclusion}: We determined the weak and strong coupling descriptions of spin-1/2 fermionic Rydberg atoms in 1D.
The various phases and transitions identified here can be probed experimentally by measuring spin-spin correlation functions with single site microscopes~\cite{Gross2021} or magnetic Bragg scattering~\cite{hart2015observation} in ultracold gases of $^6$Li.

\begin{acknowledgments}
{\it Acknowledgements}: We thank Eunmi Chae, Kaden Hazzard, and Tom Iadecola for insightful discussions. J.L. and J.H.P. are partially supported by   the Air Force Office of Scientific Research under Grant No.~FA9550-20-1-0136 and the Alfred P. Sloan Foundation through a Sloan Research Fellowship. J.H.P. acknowledges the Aspen Center for Physics, where some of this work was completed, which is supported by National Science Foundation grant PHY-1607611.
\end{acknowledgments}

\bibliography{rydberg}

\cleardoublepage

\begin{appendix}

\section*{Supplemental Material}

This supplemental material conveys additional results on the weak coupling regime, the effects of a harmonic trap, correlation functions and mutual information in the strong coupling phase separated regime, and a derivation of the spinful $PXP$ Hamiltonian at half and full filling.

\section{Weak coupling Rydberg dressed regime}

To obtain the weak-coupling Hamiltonian, see Eq. \eqref{eqn:cos} and above it, we linearize the tight-binding dispersion around $\pm k_F$. Without interactions, the bosonized Hamiltonian takes the form:
\begin{equation}
    H^{nonint}_\eta = \frac{1}{2\pi} \int dx \left[ v (\pi \Pi_\eta (x))^2 + v (\nabla \phi_\eta(x))^2 \right],
\end{equation}
where $v_F= \frac{d \varepsilon}{dk}_{k=k_F}=\frac{2t}{a}\sin(k_F a)$, where $a$ is the lattice constant of the order of $\alpha$. Next we transform the lattice interaction Eq.~\eqref{eq:rydbergint} to the momentum space:
\begin{equation}
  V_d(q) = \frac{1}{N}\sum_n 
  e^{-i q a n } V_d(a n).
\end{equation}
One can write $V_d(a n) = \int \frac{d k}{2\pi} e^{i k a n} V_C(k) = \frac{1}{L}\sum_{k=\frac{2\pi l}{L}} e^{i k a n} V_C(k)|_{L\to\infty}$, where $V_C(k) = \int dx V_d(x) e^{ikx}$ is defined in the continuum (C) limit $a\rightarrow 0$. Using the Poisson summation formula $\sum_n e^{i n g} = N \sum_m \delta_{g-2\pi m}$ one obtains:
\begin{equation}
  V_d(q) = \frac{1}{L}\sum_m V_C\left(q+\frac{2\pi m}{a}\right).
\end{equation}
As $V_C(k)$ decays as $e^{-|k|r_c/2}$ for $r_c\gg a$ the $m\neq0$ terms can be neglected and $V_d(q)\approx \frac{1}{L} V_c(q)$ used in the main text is obtained.

\begin{figure}[b!]
    \centering
    \includegraphics[width=0.9\linewidth]{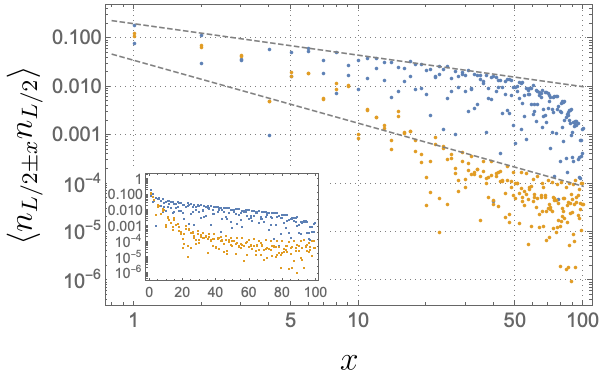}
    \caption{The density-density correlation function for $N_p = 70$ (blue) and $N_p = 120$ (yellow) particles in a $L = 200$ chain.
    All parameters are identical to that of Fig.~\ref{fig:weak}.
    The density-density correlation follows a power-law scaling on both cases. 
    }
    \label{fig:rhorho}
\end{figure}

Apart from the spin-spin correlations shown in Fig.~\ref{fig:weak}(a), we also calculate the density-density correlation 
$\langle n_{L/2 \pm x} n_{L/2} \rangle$
for the two (positive and negative backscattering) phases in Fig.~\ref{fig:rhorho}. 
As the charge sector is always gapless the density-density correlation decays as power-law in both cases.
The density-density and spin-spin power-law exponents in the repulsive backscattering region are both $-(K_\rho + K_\sigma)$. 
We extracted the exponent for $N_p = 120$ from the spin-spin correlator (dashed line in Fig.~\ref{fig:weak}(a)) and plotted a dashed line with the same slope in Fig.~\ref{fig:rhorho}. 
One can confirm that the power (slope) of the two matches well in the large $x$ limit.

\section{Effect of trap potential}

The calculations in the main text focus on an optical lattice but ignores the presence of the harmonic trap potential.
In real experiments, the harmonic trap is inevitable and will affect the previously shown results. While we expect this to have a weak effect on the weak coupling phases, we need to confirm that the phase separated regime in Fig.~\ref{fig:strong}(b) survives. We include a harmonic trap by adding to the Hamiltonian a term $\sum_{r,\sigma} V_r c_{r\sigma}^{\dag}c_{r\sigma}$ with $V_r = \frac{1}{2} m\omega^2 r^2$.

\begin{figure}[t!]
    \centering
    \includegraphics[width=0.49\linewidth]{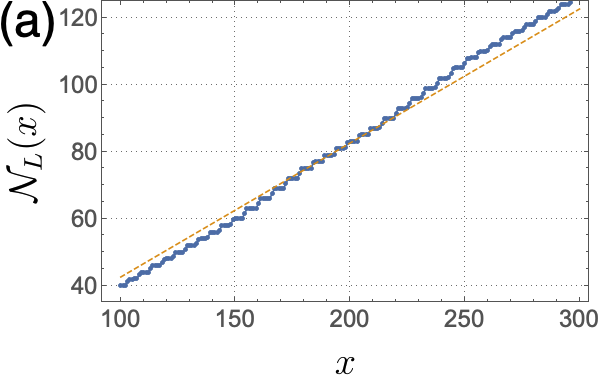}
    \includegraphics[width=0.49\linewidth]{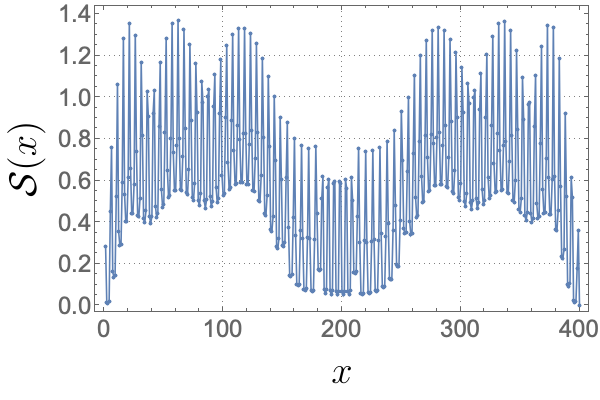}
    \includegraphics[width=0.49\linewidth]{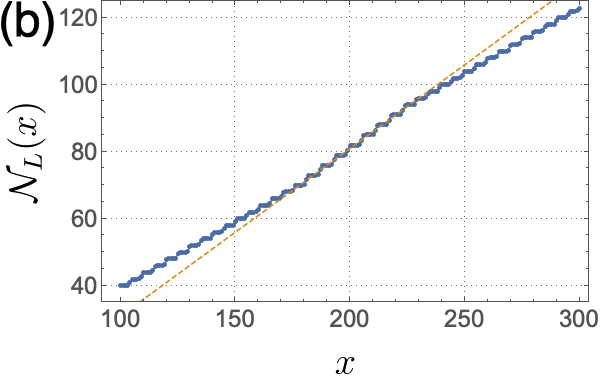}
    \includegraphics[width=0.49\linewidth]{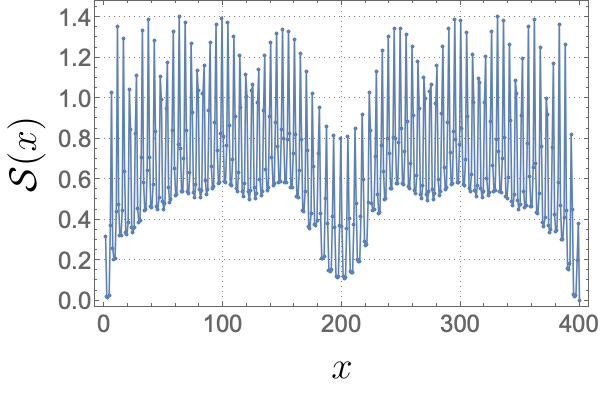}
    \caption{
    Left accumulated particle $\mathcal{N}_L(x)$ as a function of position and the bipartite entanglement entropy in a $L = 400$ chain in the presence of a harmonic trap potential.
    (a) $N_p = 167$ and $\frac{1}{2} m\omega^2 x_{\textrm max}^2 = t$, (b) $N_p = 163$ and $\frac{1}{2} m\omega^2 x_{\textrm max}^2 = 2t$.
    The yellow dashed lines are the linear fit near the center ($x = 200$) and the deviation from this line indicates different filling (and thus different phase) away from the center.
    }
    \label{fig:trap}
\end{figure}

Here, we take the experimental setup in Ref.~\onlinecite{Guardado-Sanchez2021} with $t = h \times 1.7$ kHz and $a = 752$ nm. 
Fig.~\ref{fig:trap} is the case of $N_p = 167, 163$ particles in a $L=400$ chain with a trap potential $\frac{1}{2} m\omega^2 x_{\textrm max}^2 = t, 2t$, respectively with $x_{\textrm max}=L/2$.
This corresponds to a trap frequency of order $10^2$Hz, which is safely larger than the typical harmonic trap. 

It is clear that the slope (average particle density) at the center is different from that of the two boundaries. 
The bipartite entanglement entropy also shows distinct behavior between center and boundaries, and the entanglement is low at the center regions, following the trend seen in Fig.~\ref{fig:strong}(c). Here, the center region is the magnetic domain wall and the charge like puddles are to its side. 
From these results, while the overall structure of the phase separated state is modified by the trap 
we conclude that the phase separated regime of the model will not be prohibited by the harmonic trap in experiments and we expect can be directly measured.

\section{Correlations in the phase separated regime}

\begin{figure*}[t!]
    \centering
    \includegraphics[width=0.32\linewidth]{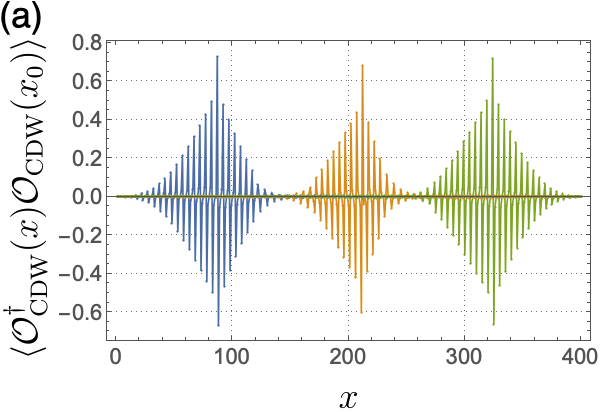}
    \includegraphics[width=0.325\linewidth]{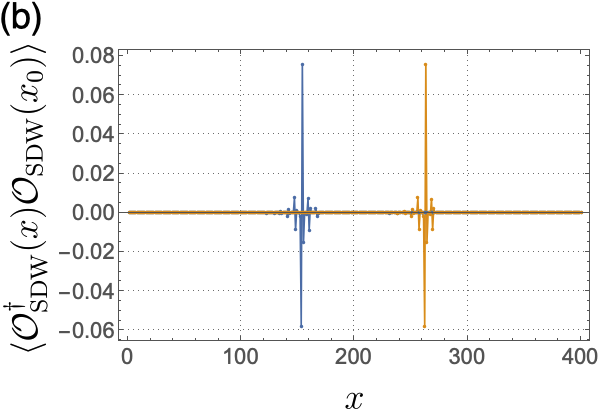}
    \includegraphics[width=0.30\linewidth]{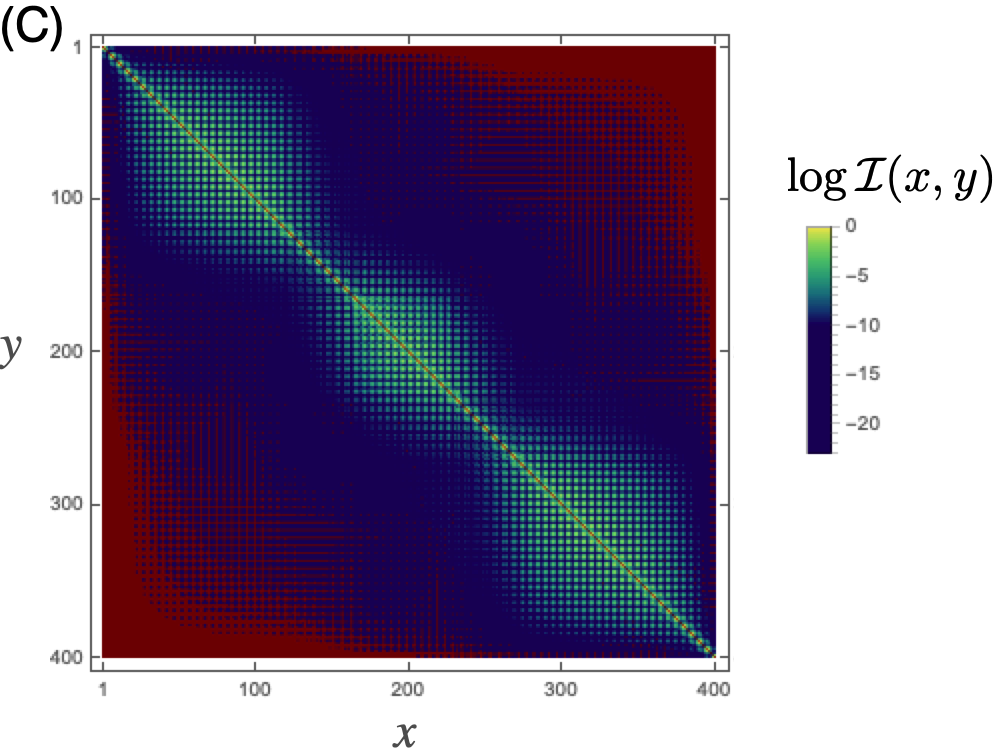}
    \caption{
    Correlation functions for the same parameters in Fig.~\ref{fig:strong}(b,c). 
    ($L=400$, $N_p=170$ in the strong coupling regime) 
    (a) The density-density correlation function where $x_0 = 87$ [blue], $212$ [yellow], $324$ [green] which are the centers of the $c=1$ puddles in Fig.~\ref{fig:strong}(c).
    (b) The spin-spin correlation function where $x_0 = 154$ [blue], $263$ [yellow] which are in between the $c=1$ puddles.
    (c) The mutual information (MI) between two points $x$, $y$ in the system.
    The MI shows three puddles which corresponds to the three $c=1$ regions, and the information spread is blocked by the magnetized region. 
    }
    \label{fig:MI}
\end{figure*}

In this section, we further investigate the phase separated state in the context of correlations to provide additional results that imply the presence of Luttinger liquid puddles.
In the main text Fig.~\ref{fig:strong}(c), we claimed that the three regions with $\bullet$$\bullet$$\circ$$\circ$$\circ$ patterns have a central charge of $c=1$ which results from the gapless charge degrees of freedom and is inherited from the weak coupling limit of the problem. 
We check this directly calculating the density-density correlation function and observing its functional form. 

In Fig.~\ref{fig:MI}(a) we plot the density-density correlation function for the same parameters in Fig.~\ref{fig:strong}(b,c), where the center $x_0$ is chosen as the center of the three $c=1$ regions. 
One can observe slow power-law decay of the correlation, although the exponent is hard to extract due to the small size of each region.
In contrast, the spin-spin correlation function decays exponentially, even in the magnetized region. 
In Fig.~\ref{fig:MI}(b) shows the fast decay of such correlation, where $x_0$ is the center of the magnetized region (in between the $c=1$ regions).

Finally, we plot the mutual information (MI) between two points $x$, $y$ in the system. 
The MI between two subsets $A$ and $B$ of the system is defined as~\cite{Amico08}:
\begin{align}
    \mathcal{I}(A, B) = \mathcal{S}(A) + \mathcal{S}(B) - \mathcal{S}(A\cup B),
\end{align}
where $\mathcal{S}(A)$ is the entanglement entropy (or von Neumann entropy) introduced in the main text.
Fig.~\ref{fig:MI}(c) plots the MI between all pair of points in the system, and shows that the MI is large only between points within the same $c=1$ region. 
This shows the natural result that mutual information is large in gapless regions, and also the interesting feature that the gapped regions serve as a barrier of information spread between the gapless $c=1$ puddles.

\section{Bosonization analysis at strong coupling}

\begin{figure}[t!]
    \centering
    \includegraphics[width=0.49\linewidth]{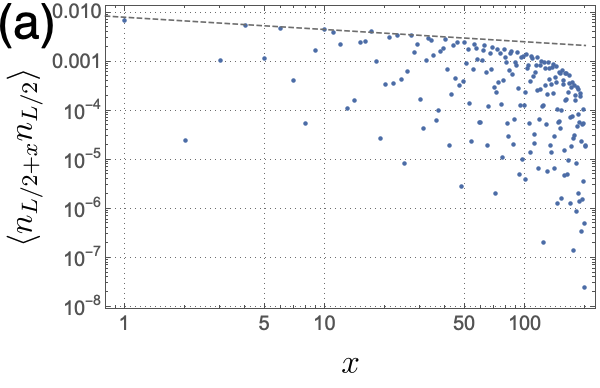}
    \includegraphics[width=0.49\linewidth]{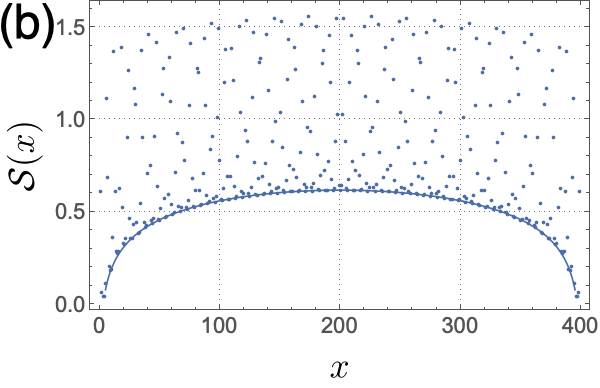}
    \caption{
    (a) The power-law scaling density-density correlation for the system in Fig.~\ref{fig:strong}(a), $N_p = 140$. (b) The bipartite entanglement entropy as a function of the cut position. The solid line is the CFT scaling form (Eq.~\eqref{eq:ee}) with $c=1$. 
    }
    \label{fig:strong2}
\end{figure}

We now present details of the bosonization discussion regarding phases for arbitrary commensurate filling. For a filling equal to $\frac{p}{q}$, $p,q$ being mutually prime integers, an additional (umklapp) term appears in the low-energy theory, corresponding to simultaneous scattering of $q$ fermions from $-k_F$ to $k_F$. This can be written as \cite{schulz_mott} 
\begin{equation}
    H_{um}=U_{um}\int dx \cos(q\sqrt{2} \varphi_\rho) \cos^q(\sqrt{2}\varphi_\sigma)
\end{equation}, where for weak coupling $U_{um}\sim U (U/t)^{q-2}$ \cite{schulz_mott,Giamarchi}. For even $q$, the most relevant part of $\cos^q(\sqrt{2}\varphi_\sigma)$ is just a constant, such that $U_{um}$ is relevant for $K_\rho<4/q^2$, leading to a charge gap in that case. Apart from half filling, this implies that the density degree of freedom can remain gapless below a threshold value of $U$. For odd $q$, $\cos(\sqrt{2}\varphi_\sigma)$ has to be kept. At weak coupling, if $V_C(2k_F)<0$ this term is fixed $\cos(\sqrt{2}\varphi_\sigma)=\pm1$ and the same condition for $K_\rho$ is valid , while for $V_C(2k_F)>0$ one can use the fixed-point value $K_\sigma=1$ \cite{schulz_mott} such that $K_\rho< 3/q^2$ is required. In the latter case, a gap will open simultaneously in the density and spin excitations. The regions of $1/5$ filling of Fig. \ref{fig:strong} (b,c) actually conform to weak coupling expectations. In that case $k_F r_c\approx 5$ such that $V_C(2k_F)/V_0 \approx - 0.16$ and a spin gap is expected, while the density gap requires strong interactions $K_\rho<0.16$. Note that the observed strong modulation of density occurs at $2k_F$, same as for a weak-coupling density wave. The 
$\odot$%
$\odot$%
$\circ$%
$\circ$%
$\circ$%
$\circ$ regions, on the other hand, has filling $1/4$ with a less stringent requirement for the charge gap $K_\rho<0.25$, allowing to understand the development of a charge gap there. While in weak coupling, a spin gap would have been also expected due to $V_C(2k_F)/V_0\approx-0.026$, its absence points to inapplicability of the continuum weak coupling analysis for $1/4$ filling here, possibly due to extremely small value of  $V_C(2k_F)/V_0$ and finite size effects. The above arguments also suggest that for 1/6 filling of Fig.~\ref{fig:strong}(a), there will be a spin gap $V_C(2k_F)/V_0\approx-0.2$ and no charge gap ($K_\rho<0.11$ is required), consistent with the numerical result. The charge modulation has the period $6$ (see Fig.~\ref{fig:strong2}(a)), corresponding to the usual $2k_F=2\pi/6$ Luttinger liquid correlations \cite{Giamarchi}, in contrast to the period $3$ ($4k_F$) modulation expected in the insulating state \cite{schulz_mott,Giamarchi}.
The bipartite entanglement entropy (see Fig.~\ref{fig:strong2}(b)) also shows clear $c=1$ scaling.

\section{Optical tweezers and the $PXP$ limit}

In the limit of no tunneling (i.e. focusing on the optical tweezer set up as opposed to an optical lattice) and strong coupling we determine the effective many-body Hamiltonian, recovering the $PXP$ model in the limit of a fully filled Fermi sea and identify a magnetic $PXP$ model with a direct exchange interaction in the limit of half-filling.

The $PXP$ limit assumes a strong mixing of ground and Rydberg state $\Omega \gg \Delta$. In this case, the dominant interaction is the repulsion between adjacent Rydberg atoms, that is also larger than $\Omega$. Essentially, the interaction forbids atoms in Rydberg state to occupy neighboring sites. For spinless bosons, Rydberg Hamiltonian reduces then to the $PXP$ model, $H = \sum_i P_{i-1}X_i P_{i+1}$~\cite{sun2008,olmos2012,turner2018}. 
$P_i = (1-Z_i)/2$ is a projector to the ground state. 
The typical $PXP$ model studied intensively in the context of quantum many-body scars and slow dynamics focused on bosonic Rydberg atoms.
Here, we consider fermionic Rydberg atoms which introduces the spin degrees of freedom to the original problem. 

First, consider a chain of Rydberg atoms with two particles (spin up and down) per site. 
The $X_i$ operator connects the $|g\uparrow, g\downarrow\rangle$ state with the $(|R\uparrow,g\downarrow\rangle + |g\uparrow,R\downarrow\rangle)/\sqrt{2}$ state and thus this model reduces to the original $PXP$ model.
This is reasonable in the sense that each site has two fermions which can be considered as a bosonic degree of freedom. 

Next, we consider a Rydberg atom chain with the filling of one particle per site. Neglecting the coupling of the laser to the spin degree of freedom (which is appropriate for $s$-like states), the Hamiltonian reduces to the original $PXP$ model with additional spin degeneracy, i.e.
\begin{align}
    H_{PXP}^{spin} = \sum_i P_{i-1} X_i P_{i+1},
\end{align}
where $X = | R \uparrow\rangle\langle g\uparrow| + | R \downarrow\rangle\langle g\downarrow| + \textrm{h.c.}$ is the Pauli matrix in the $eg$-space, $P = | g \uparrow\rangle\langle g\uparrow| + | g \downarrow\rangle\langle g\downarrow|$ is the projection to the ground state.

The spin degeneracy, however, can be lifted by interactions due to the exchange process. 
Both exchange and supexchange interactions depend on wavefunction overlap \cite{auerbach2012interacting} and can be of the same order. Therefore, one can expect that: (1) they decrease with distance fast (2) they are larger for the Rydberg state.
As the Rydberg interaction energy prohibits nearest neighbor atoms to be both in the Rydberg state, the next possibility is to either have spin-spin interactions with a neighboring ground-state atom, or a next-neighbor Rydberg atom. Combining both contributions, the spin exchange term can be written as:
\begin{equation}
\begin{gathered}
    H_{exch}^{spin} = \sum_i  J_{Rg} \vec{S}_i \cdot \vec{S}_{i+1} (P_i Q_{i+1} + Q_i P_{i+1}) 
    \\
    + J_{RR} \vec{S}_i \cdot \vec{S}_{i+2} Q_i Q_{i+2},
    \end{gathered}
\end{equation}
where $Q = 1 - P$.

For the direct exchange mechanism one can provide the following estimates for the couplings $J_{Rg}$ and $J_{RR}$. The Rydberg-Rydberg exchange coupling takes the form\cite{auerbach2012interacting} :
\begin{equation}
\begin{gathered}
     J_{RR} = \frac{1}{2} \int d^3 {\bf r} d^3{\bf r}' V_R(|{\bf r} - {\bf r}'|) \psi^*_R({\bf r}) \psi_R({\bf r}')\times 
     \\
     \times\psi_R^*({\bf r}'+2 a \hat{x} \bf) \psi_R({\bf r}+2 a \hat{x} \bf),
     \end{gathered}
\end{equation}
where $\hat{x}$ is a unit vector along the chain direction and $V_R(|{\bf r} - {\bf r}'|) \propto 1/|{\bf r} - {\bf r}'|^6$, and $\psi_{g,R}({\bf r})$ correspond to ground and excited state atomic wavefunctions. One observes then that the integral is dominated by ${\bf r} \approx {\bf r}'$ and is determined by the overlap between Rydberg wavefunctions displaced by $2a$. As the size of the Rydberg wavefunction scales with principal quantum number as $n^2$, the integral can be estimated to scale (taking the integrand value in the middle between two atoms) as $\sim \exp\left( - 2 \frac{2a}{a_B^* n^2}\right)$, where $a_B^*$ is the Bohr radius for the atom. 

The Rydberg-ground exchange coupling is determined by:
\begin{equation}
\begin{gathered}
     J_{Rg} = \frac{1}{2} \int d^3 {\bf r} d^3{\bf r}' V_{Rg}(|{\bf r} - {\bf r}'|) \psi^*_R({\bf r}) \psi_R({\bf r}')\times
     \\
     \times\psi_g^*({\bf r}'+a \hat{x} \bf) \psi_g({\bf r}+ a \hat{x} \bf),
     \end{gathered}
\end{equation}
where $V_{Rg}(|{\bf r} - {\bf r}'|)$ is the interaction between Rydberg and ground state atom, that is likely to be much weaker than the Rydberg interaction. The scaling of this integral with $n$ can be estimated analogously to be $\sim \exp\left( - \frac{a}{a_B^* n^2}- \frac{a}{a_B^*}\right)$. 
Unfortunately,  in typical tweezer setups \cite{bernien2017} $a\gg a_B^* n^2$, suggesting that the exchange interactions are likely quite small for the current set ups.

\end{appendix}

\end{document}